\documentclass[aps,showpacs,prb,amsmath,amssymb,twocolumn,superscriptaddress]{revtex4}
\usepackage{subfigure}
\usepackage{amsmath,amssymb,graphicx,hyperref}
\newcommand{\be}{\begin{equation}}
\newcommand{\ee}{\end{equation}}

\begin{document}
\title{Interaction-Enhanced Coherence Between Two-Dimensional Dirac Layers}

\author{Inti Sodemann}
\affiliation{Department of Physics, University of Texas at Austin, Austin TX 78712 USA}
\affiliation{Kavli Institute for Theoretical Physics, Santa Barbara CA 93106 USA}
\author{D. A. Pesin}
\affiliation{Department of Physics, University of Texas at Austin, Austin TX 78712 USA}
\author{A. H. MacDonald}
\affiliation{Department of Physics, University of Texas at Austin, Austin TX 78712 USA}

\date{\today}

\begin{abstract}
We estimate the strength of interaction-enhanced coherence between two
graphene or topological insulator surface-state layers by solving imaginary-axis gap equations in the random phase
approximation. Using a self-consistent treatment of dynamic screening of Coulomb interactions in the gapped phase, we show that the excitonic gap can reach values on the order of the Fermi energy at strong interactions. The gap is
discontinuous as a function of interlayer separation and effective fine structure constant, revealing a first order phase transition between effectively incoherent and interlayer coherent phases. To achieve the regime of strong coherence the interlayer separation must be smaller than the Fermi wavelength, and the extrinsic screening of the medium embedding the Dirac layers must be negligible. In the case of a graphene double-layer we comment on the supportive role of the remote $\pi$-bands neglected in the two-band Dirac model. \end{abstract}


\maketitle

\section{Introduction}

States in which quantum coherence is spontaneously established between electrons in two different layers are known as bilayer exciton condensates.   This type of ordered electronic state~\cite{naturereview} has so far been realized only in GaAs/AlGaAs double quantum wells, only at low temperatures, and even then only in the presence of a magnetic field~\cite{butov}.  When the separate layers of the bilayer condensate are contacted independently, these systems exhibit novel low bias-voltage
collective transport anomalies~\cite{condensateexperiments} related to interaction-enhanced interlayer coherence.  These anomalies could enable low-power logic devices based on new physical principles, for
example the BiSFET~\cite{bisfet} device.  It is therefore of interest to search for bilayer systems in which there is a potential for spontaneous or strongly enhanced coherence in the absence of a magnetic field
and at as high a temperature as possible.

In the absence of a magnetic field, bilayer exciton condensates are likely to form when the electron and hole Fermi surfaces in the two layers are nested. For this reason the prospects for realizing bilayer exciton condensates in the absence of a field have improved recently with the discovery of two new classes of Dirac two-dimensional electron systems, one based on graphene sheets~\cite{graphenereviews} and the other based on topological insulators~\cite{TIreviews}. In Dirac two-dimensional electron systems, electrons and holes have identical Fermi surfaces at equal carrier densities because of the linear band crossing at the Dirac point. The independent control over the carrier type and density on each layer, easily realized with gate voltages, makes these materials attractive candidates to observe interaction induced interlayer coherence.

The temperature at which the condensation sets in is difficult to estimate theoretically because, among other issues,
it is sensitive to dynamic screening of the electron-electron interaction. This problem has been studied previously using a variety of approaches and approximations~\cite{hongki,lozovik,zhang,Kharitonov,lozovik2}.
To our knowledge only one study has accounted for the dynamic nature of screening~\cite{lozovik2},
and none have accounted simultaneously for the reduction of screening~\cite{usarXiv}
that accompanies the appearance of an interlayer coherence gap.

In the present paper we show that both the dynamic nature of screening and its reduction in the coherent phase  play an essential role in determining the properties of the exciton condensate in double-layer Dirac systems. Our findings suggest that spontaneous coherence is possible at relatively high temperatures in suitable double-layer graphene systems.
(We refer to a system with two weakly hybridized graphene layers as a graphene double-layer
in order to distinguish it from a Bernal stacked two-layer graphene system, which is often referred to as bilayer
graphene.)

We use a simple approximation that employs a two-band Dirac model
for the electron and hole systems and accounts for dynamical screening
in a random phase approximation.
In single-layer graphene systems the random phase approximation
provides an accurate description~\cite{ARPEStheory} of photoemission experiments~\cite{ARPESexpt} and
in particular correctly accounts for the
plasmaron features seen near the Dirac point of systems with finite carrier densities. It can also be shown theoretically to apply in double-layer graphene systems due their large number of fermion flavors~\cite{Kharitonov}.
Here we use this approach to assess the requirements for spontaneous or enhanced coherence
in both graphene double-layer and topological insulator systems.
We conclude that strongly enhanced coherence at room temperature is possible in the graphene
case, provided that high carrier densities can still be realized at very small interlayer separations and the extrinsic dielectric screening from the substrate can largely be eliminated.
Strongly enhanced coherence at room temperature in topological insulators, on the other hand,
may require a search for new materials with larger bulk gaps.

Our paper is organized as follows.  In Sec.~\ref{model} we describe the two-band
double-layer model that we use for concrete calculations. In Sec.~\ref{screening} we discuss the properties of screening in a coherent bilayer system.
In Sec.~\ref{constantg} we analyze the interaction-enhanced coherence in a
constant gap approximation, which does not take into account the wavevector and frequency dependence of the
anomalous (electron-hole pairing) self-energy.
In Sec.~\ref{momentumg} we determine the spontaneous gap including its wave-vector dependence.  In Sec.~\ref{remotebands} we  discuss the role of the remote bands that do not intersect the Fermi surface in establishing the interlayer coherence. Finally, in Sec.~\ref{summary} we summarize our results and present our conclusions.

\section{Double-Layer Model}\label{model}

\begin{figure}
\begin{center}
\includegraphics[scale=0.8]{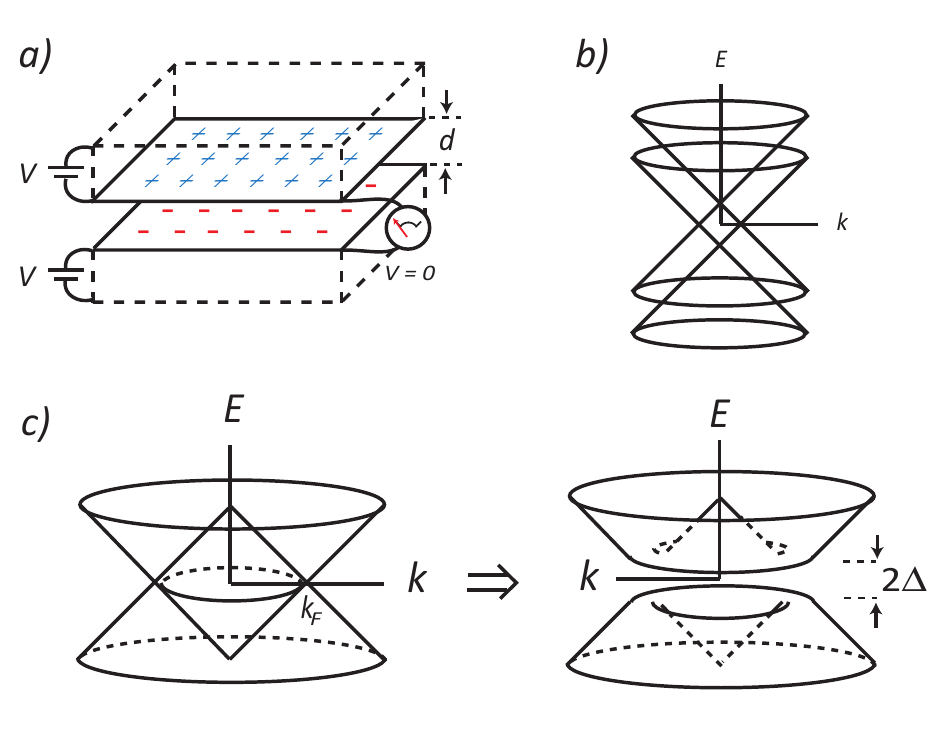}
\end{center}
\caption{\label{cones}(color online) (a) Illustration of a double-layer system. The two layers represent either two sheets of graphene separated by a thin dielectric, or the two faces of a topological insulator thin film. The layers have carriers of opposite charge and equal densities, and are in thermodynamic equilibrium. Dirac cones are displaced by the presence of carriers in each layer (b), there is
an interval of Fermi energy with only two Fermi surfaces, one from the conduction band of
the high density layer and one from the valence band of the low density layer.  For a
particular value of Fermi energy (c) the electron and hole carrier densities are equal so that
the two layers have a common Fermi surface.  Under these circumstances the
system is unstable to a broken symmetry state with spontaneous interlayer coherence and
a gap (d) in the quasiparticle spectrum.}
\end{figure}

We consider a double-layer system with a layer separation $d$.  Each layer hosts a two-dimensional electron gas that is
described by a Dirac model with velocity $v_{D}$. 
We assume that the Fermi levels $\epsilon_F$ have been adjusted so that the
magnitude of the carrier density is the same in both layers as illustrated in Fig.~\ref{cones}.
The conduction band of the electron-doped layer and the valence band of the
hole-doped layer therefore share a common Fermi radius $k_{F}$, whereas the
remote bands (the valence band of the electron-doped layer and the conduction band of the hole-doped layer)
do not intersect the Fermi energy.

Calculations described in this paper are based on a two-band model in which we
retain only bands that have a Fermi surface; the role of remote bands is discussed for both
topological insulator and graphene double-layer cases in Sec.~\ref{remotebands}. In order to apply the two-band
model consistently we impose an ultraviolet momentum cut-off at $k=2 k_F$.

We allow for the possibility that the bare Coulomb interaction strength
is reduced by a finite fraction $\epsilon$ due to extrinsic dielectric screening from the surrounding material.
(A more elaborate model is required to account for the screening from the bulk of the TI as we discuss later.)
Since both the band and interaction energies scale with distance $L^{-1}$,
this model has only three parameters when all energies are expressed
in terms of the Fermi energy $\epsilon_{F}$ : i) the Coulomb interaction strength parameter
$\alpha = e^2/\epsilon \hbar v_{D}$  ii) the Dirac band degeneracy $N$ and iii) the
dimensionless layer separation parameter $k_{F} d$.  Our main goal is to estimate how the spontaneous coherence gap at zero temperature depends on these three parameters.

The starting point of any many-body theory is a derivation of the two-particle
matrix elements of the electron-electron interaction between pairs of band eigenstates.
We allow for the presence of a non-zero bare single particle tunneling amplitude, $t_0$,
which will induce some degree of interlayer coherence even when the interactions are
absent.  Indeed, the interlayer tunneling transport anomalies on which
BiSFET operation depends require at least a modest value for $t_0$~\cite{bisfet}.
We will be interested in the many-body-enhanced tunneling amplitude, $t$,
which includes a contribution from the electron self-energy that is off-diagonal in layer index.
Spontaneous coherence occurs when $t$ remains finite for $t_0 \to 0$.
Including $t_0$, the many-electron Hamiltonian is

\begin{widetext}
\be\label{eq:hamiltonian}
H=\sum_{k,\alpha} E^0_k (c^\dagger_{k\alpha} c_{k\alpha}-v^\dagger_{k\alpha} v_{k\alpha}) +\frac{1}{2 A}\sum_{k_i,X,X'} \cos\Bigl(\frac{\phi_1-\phi_4}{2}\Bigr) \cos\Bigl(\frac{\phi_2-\phi_3}{2}\Bigr)V^0_{X,X'}(k_2-k_3)b^\dagger_{Xk_4\alpha}b^\dagger_{X'k_3\beta}b_{X'k_2\beta}b_{Xk_1\alpha} \; \delta_{k_1+k_2,k_3+k_4}
\ee
\end{widetext}

where $\{b_{Xk\alpha}^{\dagger},b_{Xk\alpha}\}$ are the creation and annihilation operators
for Dirac band states with definite two-dimensional momentum $k$ and layer index $X$,
while $\{c^\dagger_{k\alpha},c_{k\alpha}\}$ and $\{v_{k\alpha}^{\dagger},v_{k\alpha}\}$
are creation and annihilation operators for the respective conduction and valence bands in the
presence of the interlayer hybridization amplitude $t_0$.  In this equation
$E^0_k = \sqrt{v_{D}^2 (k-k_{F})^2 + t_0^2} $ is the bare band energy magnitude,
$\phi_i$ is the orientation angle for momentum $k_i$,
and  $k_{F}$ is the common Fermi momentum of conduction and valence bands in the absence of
hybridization.  The Greek indices
on the creation and annihilation operators range over the number $N$ of Dirac fermion flavors.
When the Dirac model is viewed as a continuum model of graphene, we should choose $N=4$ to
account for two valleys and two spin states, whereas for the surface states of topological insulators like Bi$_2$Se$_3$
we have $N=1$~\cite{TIreviews}.

The linear dependence of band energy on momentum for $t_0=0$
in Hamiltonian~(\ref{eq:hamiltonian}) is a hallmark of the Dirac physics of
graphene and topological insulators.
The interaction term describes pairs of electrons which are scattered by their mutual interactions.
In this term $V^0_{X,X'}(q)$ is the two-dimensional Fourier transform of $e^2/r$ which equals
$V_{S}(q)= 2\pi e^2/q$ for electrons in the same layer and $V_{D}(q) = \exp(-qd) V_{S}(q)$
for electrons in different layers.
The factors of $\cos((\phi_1-\phi_4)/2)$ and $\cos((\phi_2-\phi_3)/2)$ which appear in this equation are well
known in the graphene literature, and are due to the dependence of the phase difference
between sublattice components of the band states on the direction of momentum.

\section{Bilayer Screening}\label{screening}

\begin{figure}
\begin{center}
\includegraphics[scale=0.9]{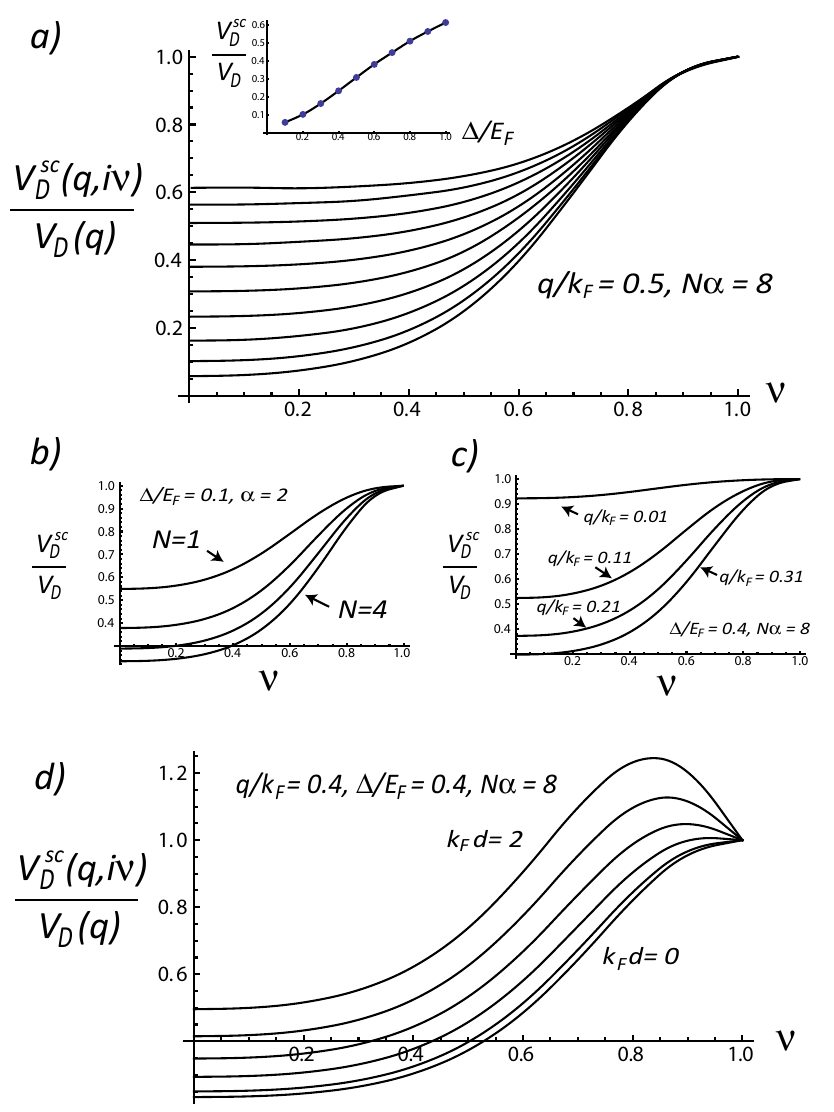}
\end{center}
\caption{Ratio of screened interlayer potential to bare interlayer potential as a function frequency $\omega$ along the imaginary axis.  In these plots $\nu$ is a non-linear dimensionless frequency defined by $\omega/\epsilon_F\equiv \nu/(1-\nu)$ so that $\omega=\epsilon_F \Leftrightarrow \nu=1/2$ and $\omega=\infty \Leftrightarrow \nu=1$.  Screening power increases with the product of the interaction strength $\alpha$ and the number of Dirac flavors $N$ in each layer.  The value $N\alpha = 8$ corresponds to
graphene surrounded by low $\epsilon \sim 1$ dielectric material.  Screening weakens at larger frequencies and larger gaps, vanishing in the high frequency limit as illustrated in panel (a) where the curves correspond to gaps $\Delta/\epsilon_F=\{0.1,0.2,...,1.0\}$ (The inset shows the zero frequency value of this ratio as function of the gap.).
The flavor number $N$ enhances screening as illustrated in panel (b).
In isolation this consideration would favor enhanced coherence in the $N=1$ topological insulator thin film case compared to the
$N= 4$ graphene double-layer case.
Screening vanishes for $q,\omega \to 0$ as illustrated in (c). The screened interaction
can sometimes be stronger than the unscreened interaction as illustrated in panel (d), where the curves
 correspond to dimensionless layer separations $k_F d=\{0,0.4,0.8,...,2.0\}$, because of the
interplay of even channel and odd channel screening explained in the text.
}
\end{figure}

In a bilayer, Coulomb interactions in one layer induce a charge response in the same layer and in
the opposite layer.   We use the random phase approximation for screening in which electrons respond
like non-interacting particles to the sum of the external potentials and the mean-field
Hartree potentials from the charge densities that they induce.
The dynamic non-interacting response functions can be calculated by using first order time-dependent
perturbation theory.  When the responding electrons are coherent combinations of the two layers
we find that the imaginary frequency same-layer and different-layer single-particle
density response functions are:
\begin{widetext}
\be
\Pi_S(q,i\omega)=-N\int\frac{d^2 k}{(2\pi)^2}\frac{1+\cos(\phi_k-\phi_{k+q})}{2}\Bigl(1-\frac{\xi_k \xi_{k+q}}{E_k E_{k+q}}\Bigr)\frac{E_k}{\omega^2+(E_k+E_{k+q})^2}.
\ee
and
\be
\Pi_D(q,i\omega)=N t^2\int\frac{d^2 k}{(2\pi)^2}\frac{1+\cos(\phi_k-\phi_{k+q})}{2}\frac{1}{E_k(\omega^2+(E_k+E_{k+q})^2)}.
\ee
\end{widetext}

where $\xi_k=v_D (k-k_{F})$, $E_k = \sqrt{v_{D}^2 (k-k_{F})^2 + t^2}$, and $t$ is understood as the full interaction enhanced hybridization amplitude.
In solving the gap equations discussed in the next section, we will find it convenient to take advantage of the smooth
frequency dependence of the polarization functions along the imaginary axis.

For an equal carrier density system, the screening of the sum $V_{+}=V_{S}+V_{D}$ (even screening channel) and difference
$V_{-} = V_{S}-V_D$ (odd screening channel) of the intralayer and interlayer interactions decouple.  An elementary calculation
yields simple geometric series expressions for the even and odd screened interactions:
\be
V_{\pm}^{sc} = \frac{V_{\pm}}{1-V_{\pm} (\Pi_{S} \pm \Pi_{D})}.
\ee
Recalling that $E_{k}^2=\xi_k^2+t^2$, it is easy to show that $\Pi_{S} + \Pi_{D}$
vanishes for $q \to 0$.  There is no even channel polarization and no even channel screening in this limit because conduction and valence quasiparticle bands at the same $k$  are orthogonal.  More fundamentally, the total density response of a gapped system to a uniform potential shift is zero.  For the $q,\omega \to 0$ limit, the odd channel perturbation on the other hand, corresponds simply to a shift in Fermi wavevector. It follows that the odd channel polarization approaches the thermodynamic density of states in this limit, $\Pi_S-\Pi_D \to -N\nu_0$ up to small corrections $\mathcal{O}(t/\epsilon_F)^2$,  and it is not substantially altered by interlayer coherence ($\nu_0=k_F/(2\pi v_D$)).  For $q,\omega \to 0$, $V_{D} \to V_{S}$ so that there is no
odd channel potential, and screening is therefore absent as illustrated in Fig. 2.   For larger $q$,
$V_{S} > V_{D}$ and the screened interlayer interaction can therefore sometimes be stronger than the bare interlayer interaction.
After a bit of simple algebra, we find the following general expression for the
random phase approximation screened interlayer interaction:
\be\label{Vsc}
V_D^{sc}=\frac{V_D+\Pi_D (V^{2}_S-V^{2}_D)}{1-2(V_S \Pi_S+V_D \Pi_D)+(V^{2}_S-V^{2}_D) (\Pi_S^2-\Pi_D^2)}.
\ee
The screened interlayer interaction is invariably larger than the value $1/(2N\nu_0)$ obtained when the
polarization functions are replaced by their zero frequency normal state values.  We therefore expect to find
stronger interlayer coherence in our calculations, which account for the retarded (frequency dependent) character
of the Dirac gas polarization and include the effects of the gap in screening, than in the estimates of Ref.~\onlinecite{Kharitonov} in which a static screening approximation for the polarizablizities in the normal phase was employed.

\section{Constant Gap Approximation}\label{constantg}

In many-body theory~\cite{vignale,mahan} the influence of electron-electron interactions on electronic structure is
captured by the self-energy.  The self-energy can be thought of as an
interaction correction to the quasiparticle Hamiltonian, like the exchange and correlation corrections that
appear in density-functional theory, except that it varies with the momentum and energy of the
electron-wave under consideration.  Our analysis of interlayer hybridization is based on the
simplest sensible approximation for the self-energy in an interacting electron system.
The approximation accounts for screening, which plays an essential role in any
interacting electron problem in metals because it converts the long-range $e^2/r$ Coulomb interaction
into a shorter range interaction that has a finite value when integrated over space.
It also accounts for the time-delay of screening responses, retardation, which makes
screening less effective when inter-electron potentials fluctuate dynamically.
We approximate the self-energy by the first order term in its perturbation theory expansion
in terms of dynamically screened Coulomb interactions.  This approximation is 
commonly employed in {\em ab initio} electronic structure theory calculations where it is known as the $GW$ approximation~\cite{GWapp1,GWapp2}.  In the graphene case this approximation has been employed to study a variety of physical properties both in the context of the Dirac equation electron gas model~\cite{ARPEStheory} and in {\em ab initio}~\cite{abinitio_gw_refs} calculations and appears to be accurate where comparisons with experiment are possible.
One of the great successes of $GW$ theory more broadly is its
explanation for the property that band gaps in semiconductors are almost always larger
than predicted by density-functional-theory~\cite{GWapp1,GWapp2}.  This property is closely related to the enhanced
interlayer coherence on which we focus, which may be viewed as simply another example
of this well known tendency.  The interaction enhanced gaps depend essentially on the
non-locality of screened exchange interactions which is neglected by the
local density approximation and in most exchange-correlation approximations that
are commonly used in {\em ab initio} density-functional-theory calculations.

The recipe for constructing the $GW$ approximation for the self-energy 
from the Hamiltonian is well-established and leads in the case of a double-layer 2DEG Dirac model to a
self-energy that is wavevector and frequency dependent and is a matrix in layer index:
\begin{widetext}
\be
\Sigma_{X,X'}(q,i\Omega)=\int\frac{d^2 k}{(2\pi)^2}\frac{1+\cos(\phi_q-\phi_{k+q})}{2}\int_{-\infty}^{\infty}\frac{d\omega}{2\pi}G_{X,X'}(q+k,i(\Omega+\omega)) \; V_{X,X'}^{sc}(k,i\omega),
\ee
\end{widetext}
where the Green's function is related to the self-energy through the Dyson equation: $G=(G_0^{-1}-\Sigma)^{-1}$. We have chosen to take advantage of the analytic properties of the dependence of screened interaction on complex
frequency to evaluate this quantity along the imaginary axis, where its frequency-dependence is smoother than along the real one, and thus easier to treat numerically.

The particle-hole and layer symmetry of our model leads to a self-energy matrix of the form

\be
\Sigma(q,i\omega)=\begin{pmatrix}
K(q,i\omega)& \Delta(q,i\omega)\\
\Delta(q,i\omega)& -K^*(q,i\omega)
\end{pmatrix}.
\ee
Using the explicit form of the bare Green's function with interlayer tunneling  amplitude $t_0$,
\be
G_0^{-1}(q,i\omega)=\begin{pmatrix}
i\omega-\xi_q& -t_0\\
-t_0& i\omega+\xi_q
\end{pmatrix},
\ee
we find that the off-diagonal component of the self-energy satisfies the following equation:

\begin{widetext}
\be\label{biggapeq}
\Delta(q,i\omega')=\int\frac{d^2 k}{(2\pi)^2} \; \frac{1+\cos(\phi_{q}-\phi_{k})}{2} \,\int_{-\infty}^{\infty}\frac{d\omega}{2\pi} \;
\frac{t_0+\Delta}{(t_0+\Delta)^2+(\omega-\textrm{Im}K)^2+(\xi_k+\textrm{Re}K)^2}
\; V^{sc}_D(q-k,i(\omega'-\omega)).
\ee
\end{widetext}

Throughout this study we neglect the interaction correction to the diagonal self-energy $K(q,i\omega)$. Instead, we focus on the off-diagonal anomalous self-energy $\Delta(q,i\omega)$, whose value determines the excitonic gap. For gaps much smaller than the Fermi energy the integral kernel in Eq.~\eqref{biggapeq} is so highly peaked at the Fermi surface, that any trial gap function which only deviates from the actual one at large momenta and frequencies will not affect substantially the self-consistent determination of the value of the gap at $\omega=0$ and $q=k_{F}$. In this section we discuss the simplest self-consistent procedure which exploits this property, in which $\Delta(q,i\omega)$ is replaced by its value at $\omega=0$ and $q=k_{F}$, {\em i.e.} by its value at the center of the wavevector-energy region in which coherence is established. We will refer to this approximation as the constant gap approximation.

The main merit of this approximation is that it allows a transparent discussion of the dependence of
the coherence gap on the three dimensionless quantities, $\{\alpha, k_F d, N\}$, that parametrize the problem. The constant gap approximation is relaxed in the next section, in which the coherence gap is calculated more accurately by including its momentum dependence in the integral equation.

In this approximation the interaction-enhanced interlayer tunneling amplitude is $t=t_0+\Delta$, where
\begin{widetext}
\be\label{eq:spontaneous_gap}
\Delta=\int\frac{d^2 k}{(2\pi)^2}\frac{1+\cos(\phi_{q}-\phi_{k})}{2}\int_0^{\infty}\frac{d\omega}{\pi}\frac{t}{\omega^2 + t^2+\xi_{k}^2}
\; V^{sc}_{D}(|k - k_F \hat{q}|,i\omega) .
\ee
\end{widetext}
Here $\hat{q}$ is the direction of the wavevector on the Fermi surface for which we evaluate $\Delta$;
because of the isotropy of the problem $\Delta$ does not depend on $\hat{q}$.

\begin{figure}
\begin{center}
\includegraphics[scale=0.53]{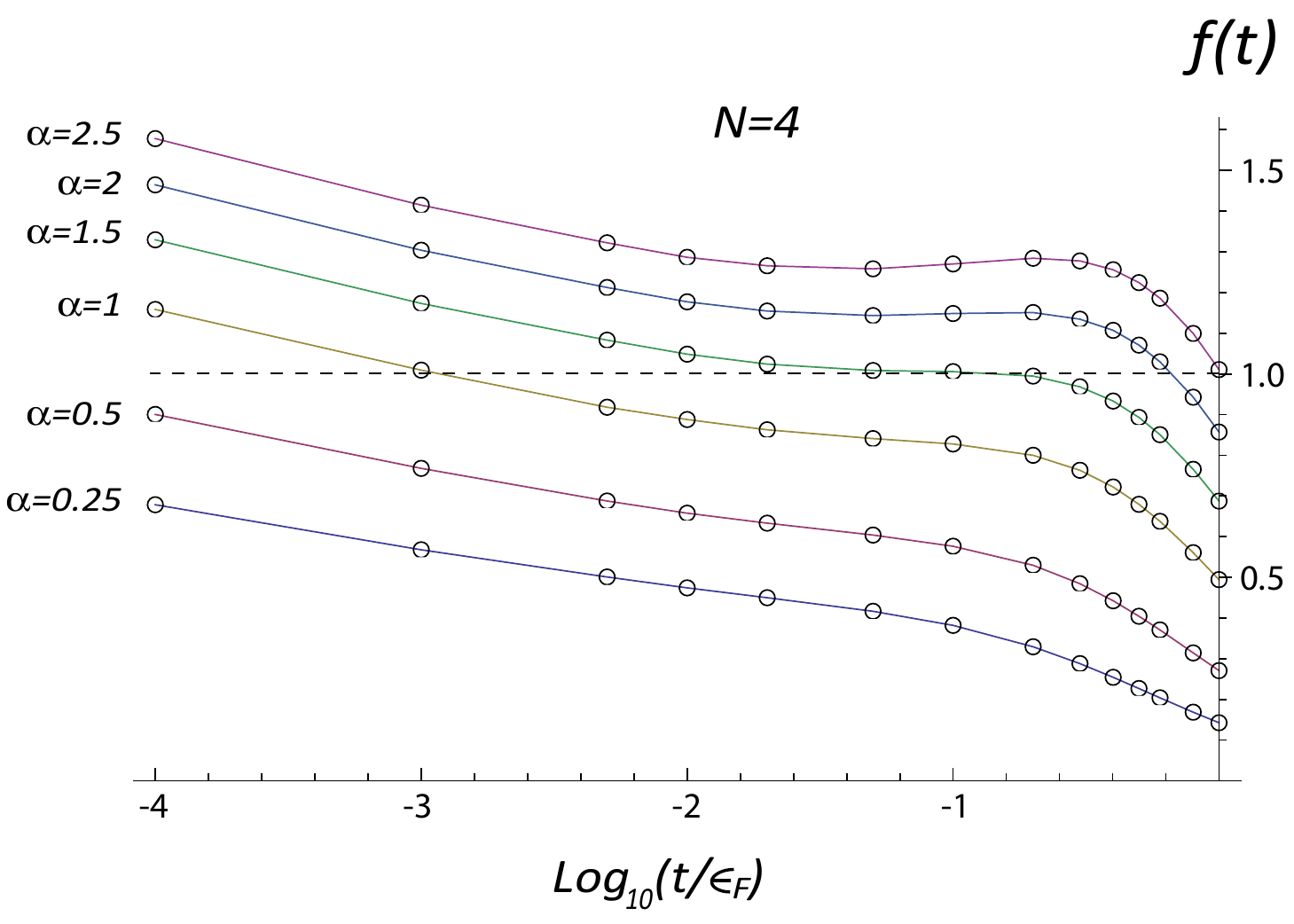}
\end{center}
\caption{\label{fvsgap} $f(t)\equiv\Delta/t$ {\em vs.} $t/\epsilon_{F}$ on a logarithmic scale for $N=4$ and layer separation $d \to 0$ at a series of values of $\alpha$. The linear dependence of $f$ on $-Log_{10}(t/\epsilon_F)$ at very small gaps is a result of the nesting of electron and hole Fermi surfaces as explained in the text. The size of the spontaneous gap is determined by solving $f(t)=1$.  This figure illustrates why the spontaneous gap can be $\sim \epsilon_F$ at strong interactions and orders of magnitude smaller at weak interactions.}
\end{figure}

Strictly speaking, bilayer exciton condensation is characterized by
the spontaneous appearance of interlayer coherence in the absence of any
interlayer hybridization.  Because single-particle processes that support
interlayer hybridization are always present, strict
exciton condensation is always a theoretical concept, but one which
is still useful in considering real physical properties.

The constant gap
approximation is conveniently discussed in terms of the quantity
\begin{widetext}
\be\label{eq:f-func}
f(t)\equiv\frac{\Delta}{t} =
\int\frac{d^2 k}{(2\pi)^2}\frac{1+\cos(\phi_{q}-\phi_{k})}{2}\int_0^{\infty}\frac{d\omega}{\pi}\frac{1}{\omega^2 + t^2+\xi_{k}^2}
\; V^{sc}_{D}(|k-k_F \hat{q}|,i\omega).
\ee
\end{widetext}
Coherence is possible in the absence of a single-particle contribution when $f(t)=1$.
This equation always has a solution since $f(t) \to 0$ for very large $t$,
but diverges logarithmically for $t \to 0$. In Fig.~\ref{fvsgap} we plot $f(t)$ vs $t$ for $N=4$ for a series of $\alpha$ values.

Equation~(\ref{eq:spontaneous_gap}) for the spontaneous gap is identical to the equation for the superconducting gap
in the BCS theory of superconductuctivity, except that in the Coulomb interaction case we must account for
dynamic screening. The logarithmic divergence at small $t$ can be traced to the nesting between conduction and valence
band Fermi surfaces in bilayers with overall neutrality. Because the numerical coefficient in front of the logarithmic divergence is strongly reduced by screening~\cite{Kharitonov}, $f(t)$ reaches $1$ only for extremely small values of
$t$ when interactions are weak.

\begin{figure}
\begin{center}
\includegraphics[scale=0.6]{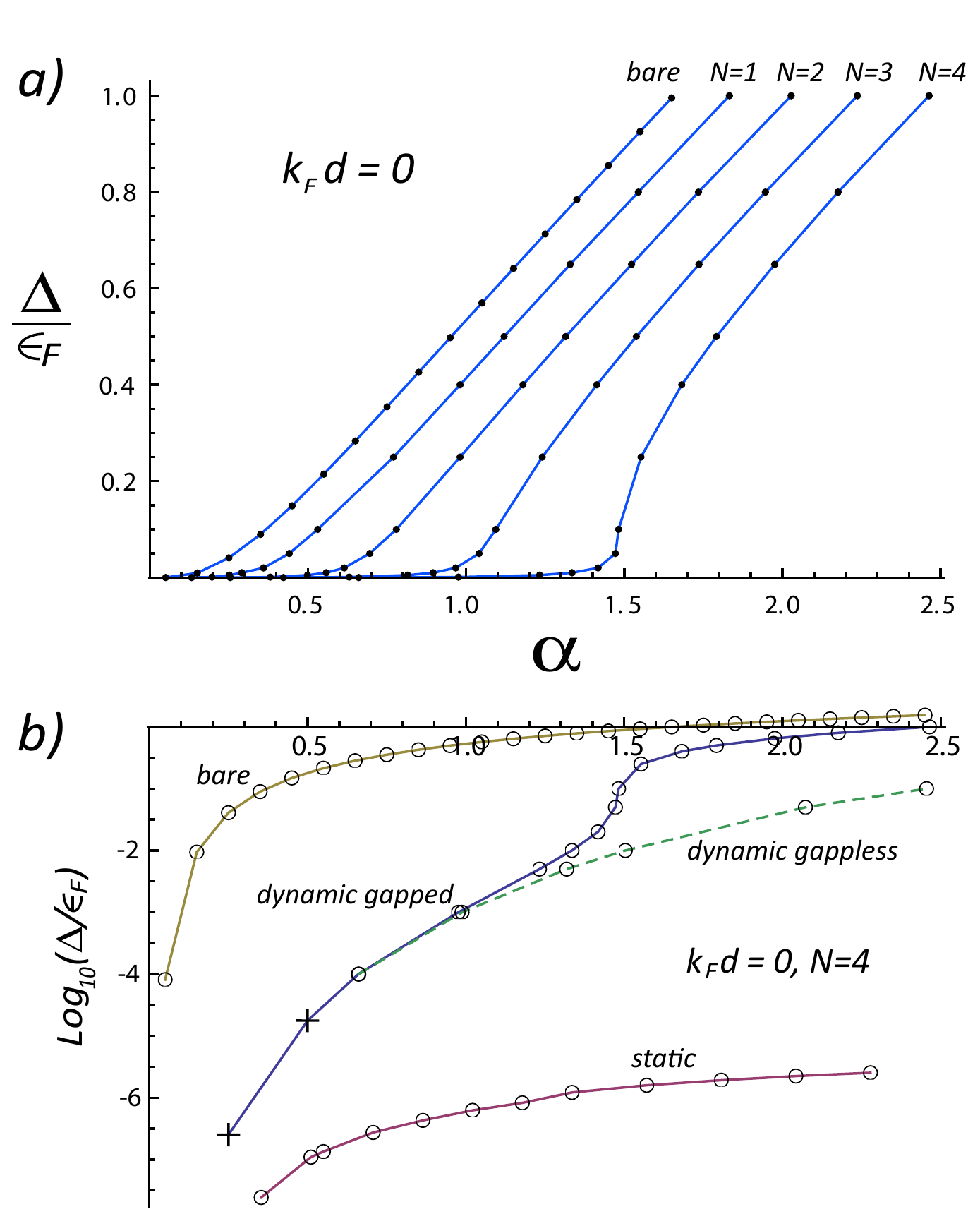}
\end{center}
\caption{\label{spont_gap}a) Constant gap approximation
spontaneous gap as a function of effective fine structure constant $\alpha=e^2/(\varepsilon v_D)$, for (left to right) unscreened interactions and $N=\{1,2,3,4\}$ degenerate Dirac cones ($N=1$ for Bi$_2$Se$_3$, $N=4$ for graphene). b) Comparison between several approaches to screening.  When the screening is treated by using static screening an extremely small gap is predicted. 
If the reduction in screening in the coherent state is neglected
the sudden rise at $\alpha\sim 1.5$ is absent (for the dashed line we fixed the gap to be $10^{-4} \epsilon_F$ inside the polarization functions, thus virtually computing them in the gapless phase.). The full dynamical and gapped screening result approaches the result obtained with bare interactions at strong coupling but differs from it by several orders of magnitude at weak coupling (the crosses in the dynamic gapped screening case were obtained by extrapolating the $f$ function to very small gaps see Fig.~\ref{fvsgap}).}
\end{figure}

The role of screening changes dramatically when interactions are strong.
In Fig.~\ref{spont_gap} we have plotted the spontaneous gap as a function of the dimensionless interaction parameter $\alpha=e^2/\varepsilon v_D$ in the limit of vanishing interlayer separation $k_F d\rightarrow 0$, which is the most favorable limit for excitonic condensation. In graphene a steep increase in the gap occurs near $\alpha \sim 1.5$, where it rises to become of the order of the Fermi energy. This result underlines the importance of the dielectric medium surrounding the double-layer having as low a dielectric constant, $\epsilon$, as possible.

\begin{figure}
\begin{center}
\includegraphics[width=3.3in]{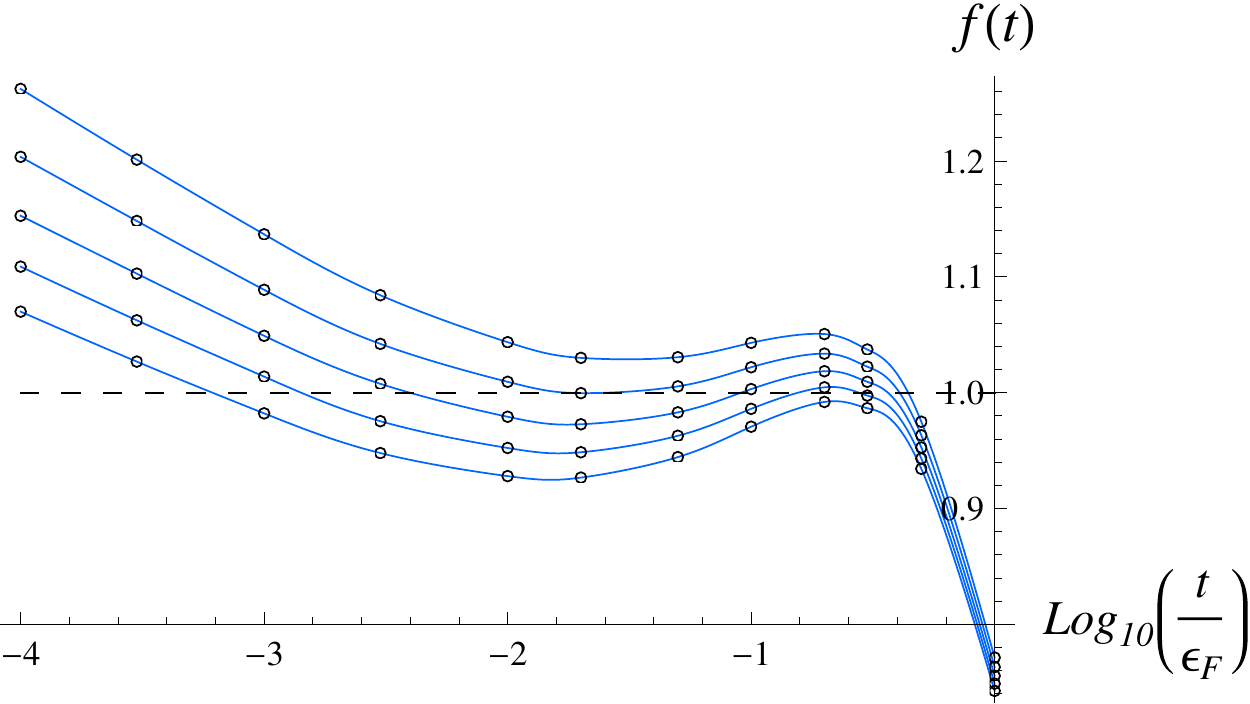}
\end{center}
\caption{\label{fig:qpt}(color online) Evolution of $f(t)$ with increasing interlayer distance. In these curves $\alpha=1.8$ and $k_F d=\{0.04,0.06,...,0.12\}$. Metastability exists for $k_F d_{\rm{cr2}}>k_Fd>k_F d_{\rm{cr1}}$, with $k_F d_{\rm{cr1}}\sim0.06$ and $k_F d_{\rm{cr2}}\sim0.11$.}
\end{figure}

Self-consistent screening of Coulomb interaction in the presence of the excitonic condensate leads to the possibility of a  first order quantum phase transition between states with very different values of the excitonic gap as the interlayer separation is changed. This originates from the tendency of the condensate to reduce screening, thus the screened interaction, $V^{sc}_D$, in Eq.~(\ref{eq:spontaneous_gap}) is an increasing function of $\Delta$. At strong interactions this can lead to non-monotonic behavior of the right hand side of Eq.~(\ref{eq:f-func}) as function of $\Delta$, as illsutrated in Fig.~\ref{fig:qpt}.  

\begin{figure}
\begin{center}
\includegraphics[scale=0.6]{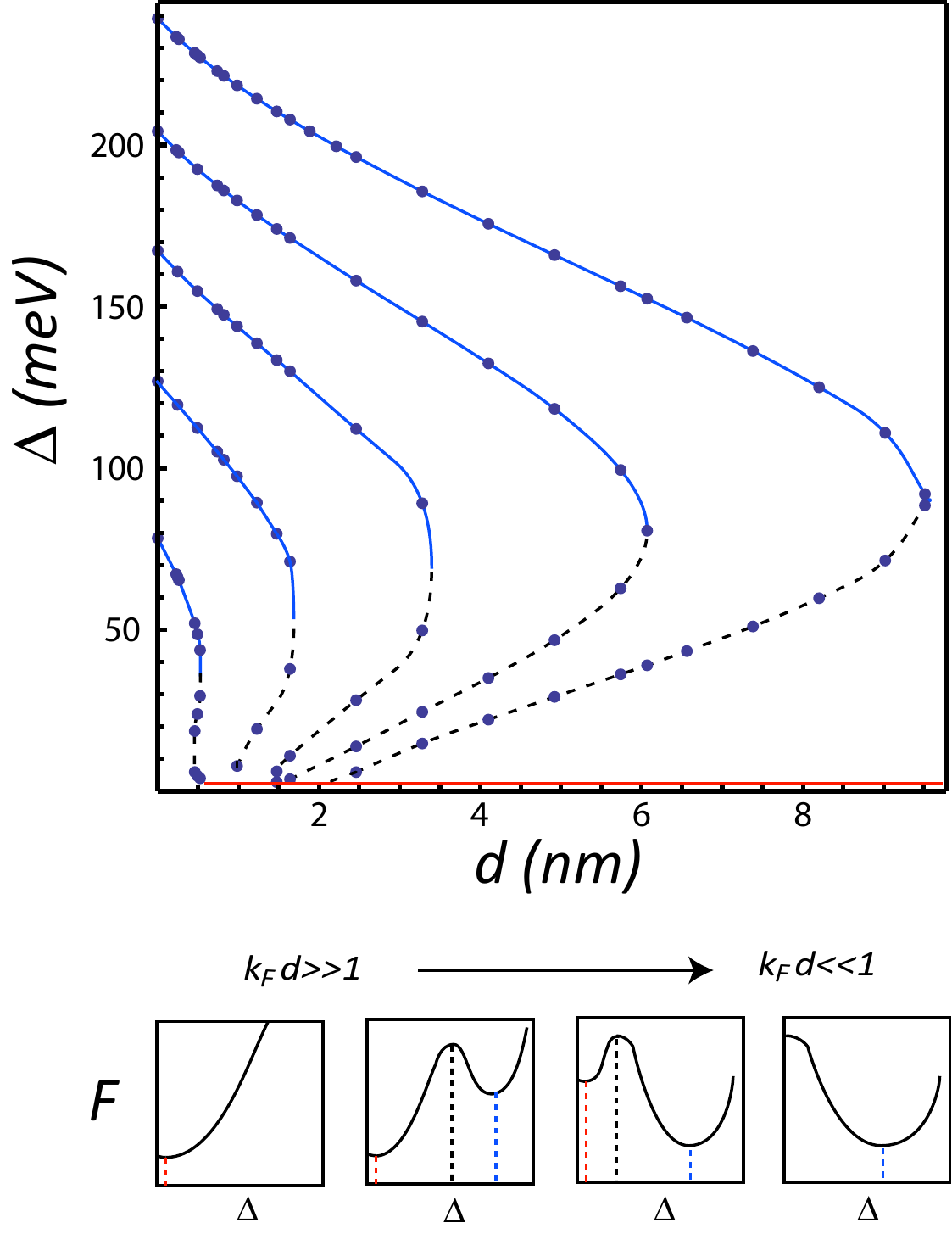}
\end{center}
\caption{\label{graphene_d}(color online) Spontaneous gap in graphene as a function of interlayer separation (with $\epsilon_F = 250$ meV) for  $\alpha=\{1.6,1.8,2.0,2.2,2.4\}$ (for graphene suspended in vacuum $\alpha\sim 2.2$, for an SiO$_2$ substrate $\alpha \sim 1.3$). As discussed in the text there can be three self-consistent solutions at a given interlayer separation, the dashed line depicts the unphysical solution, the solid blue line depicts the physical solution where the gap reaches values on the order of the Fermi energy at small interlayer distances, and the red line schematically depicts the solution where the gap is much smaller than the Fermi energy at large interlayer separations. The panels below present the schematic behavior of the free energy as
a function of the gap, illustrating the first order quantum phase transition.}
\end{figure}

When plotted as a function of interlayer distance for fixed $\alpha$, the gap is a multivalued function for a window of distances. The branch on which it appears to increase as a function of distance at fixed densities and $\alpha$ is unphysical, and the other two branches correspond to metastable states. Figure~\ref{graphene_d} illustrates this behavior for graphene with $\epsilon_F=0.25eV$ and a series of $\alpha \geq 1.6$. Thus at fixed $\alpha$ and large interlayer distances $d\gg 1/k_F$ the gap is several orders of magnitude smaller than $\epsilon_F$, this tendency persists down to zero interlayer separation for small fine structure constants, but for $\alpha \gtrsim 1.5$, a first order phase transition to a coherent state with a gap on the order of $\epsilon_F$ is encountered at a critical distance $d_{cr} \lesssim 1/k_F$. Metastability of the two states can persist down to a second critical value of the interlayer distance, after which only the state with strong interlayer coherence is a self-consistent solution of the gap equation.

\begin{figure}
\begin{center}
\includegraphics[scale=0.55]{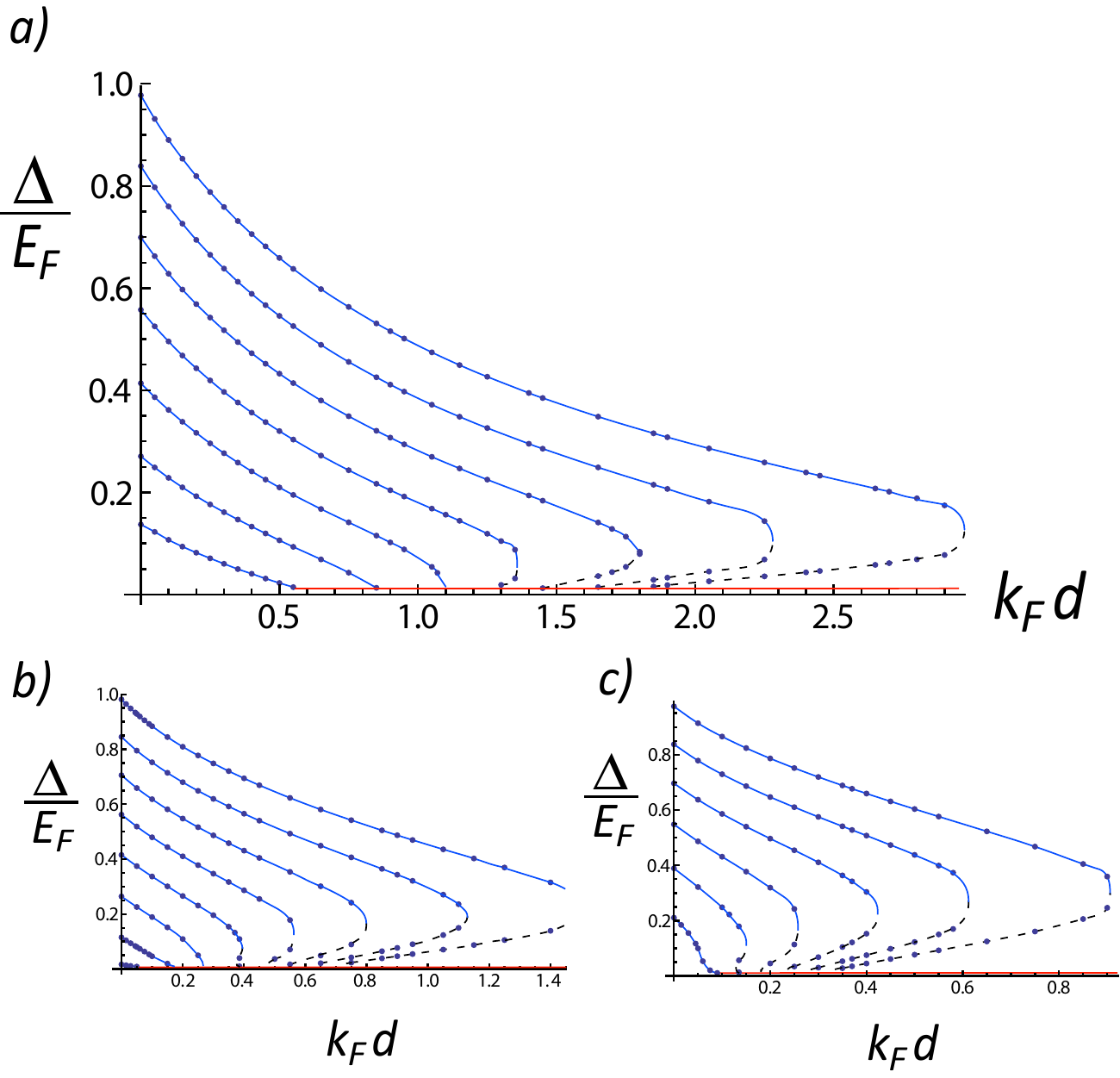}
\end{center}
\caption{\label{TI_d} Spontaneous gap for a $N=1$ topological insulator like Bi$_2$Se$_3$, as a function of the dimensionless interlayer separation $k_Fd$ for $\alpha=\{0.6,0.8,1.0,1.2,1.4,1.6,1.8\}$.
In the absence of dielectric screening
$\alpha\sim 4.4$ for Bi$_2$Se$_3$  because its Dirac velocity is smaller than that of graphene.
The interaction strength is reduced even in vacuum due to screening by the bulk of the topological insulator.
The lower panels are for $N=\{2,3\}$. The first order phase transition is present at strong interactions
but it is less pronounced as $N$ decreases.}
\end{figure}

The same qualitative behavior is found for TIs as illustrated in Fig.~\ref{TI_d}, although the first order phase transition becomes less pronounced at small $N$. Figure~\ref{TI_d} also illustrates that the ratio $\Delta/\epsilon_F$ is a decreasing function of $k_F d$, this implies that for a given device, with a fixed interlayer distance and effective fine structure constant, the excitonic gap is not expected to be a monotonic function of $\epsilon_F$, but would increase at small $\epsilon_F$, be peaked for a value $\epsilon_F \lesssim v_D/d$, and decrease at larger values.

The existence of a first order phase transition as a function of distance at zero temperature suggests the existence of a first order transition as a function of temperature at fixed $\{\alpha,k_Fd\}$, and thus a calculation where a continuous phase transition to the excitonic phase is assumed and the finite temperature linear gap equation is solved with screening computed in the normal phase would tend to underestimate the critical temperature for this transition.

Spontaneous coherence in topological insulators is favored by their reduced number of degenerate Dirac cones~\cite{moore}.
On the other hand the 3D bulk of the material is expected to substantially screen the interactions between
the surface Dirac electrons, strongly reducing their effective fine structure constant from its nominal value in vacuum
$\alpha_0=e^2/v_D$.  In addition, the maximum possible value of the Fermi energy is the energetic separation between
the Dirac point and the closest band, which is $\sim 0.1 {\rm eV}$ in current topological insulators. Even if large excitonic gaps were realized at zero temperature, the Kosterlitz-Thouless transition would limit the critical temperature to $\sim 0.1 \epsilon_F \sim 0.01 {\rm eV}$~\cite{hongki}, this alone poses a major limitation to achieve ambient temperature inter-surface coherence in currently known topological insulators.

\begin{figure}
\begin{center}
\includegraphics[scale=0.5]{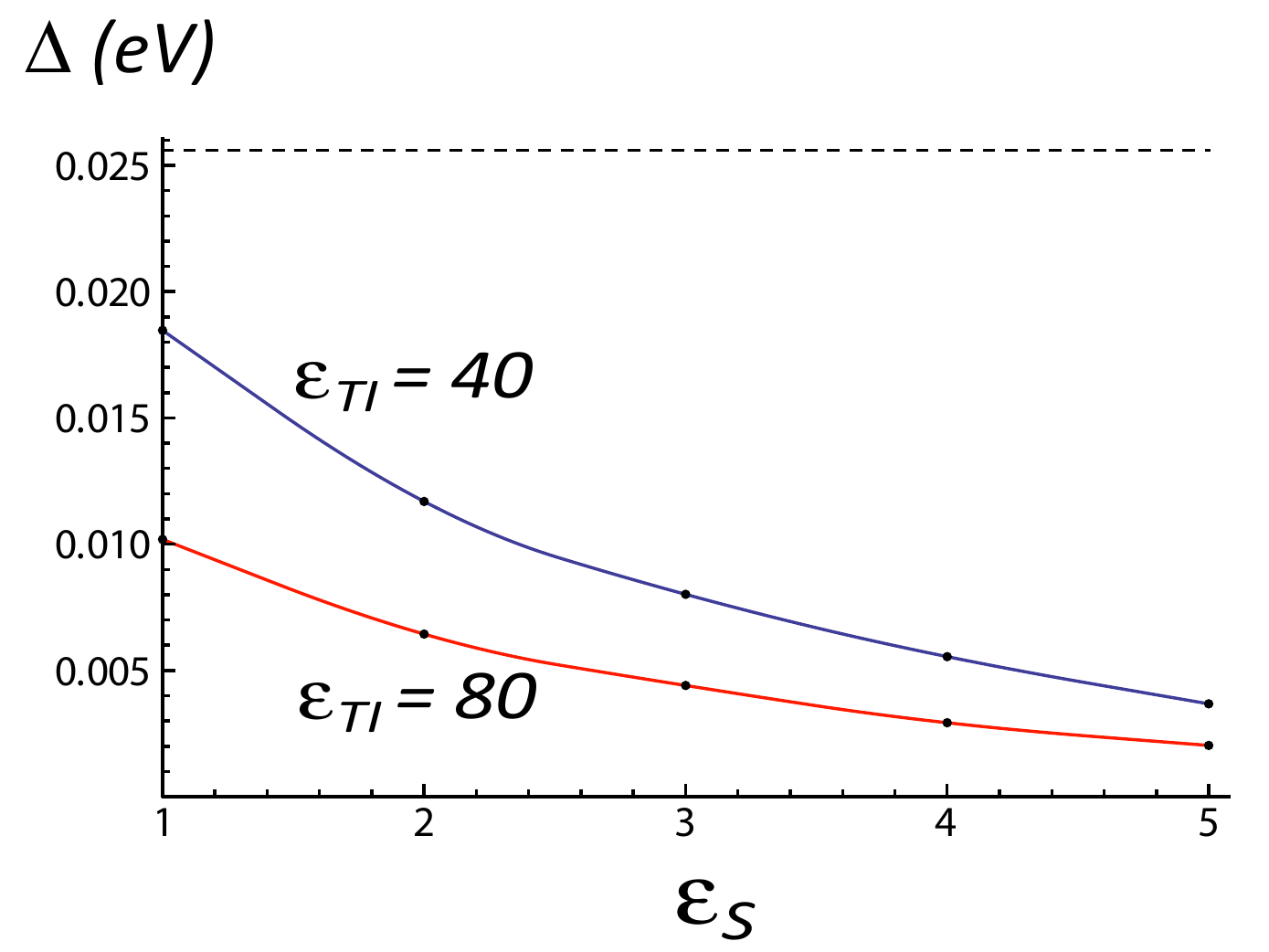}
\end{center}
\caption{\label{TIscreening} Interaction induced gap in a thin film topological insulator, with parameters suitable for Bi$_2$Se$_3$. These results correspond to film thickness of $d_0 = 3nm \sim 3QL$ and the fixed ratio $\Delta/\epsilon_{F}=0.1$. For larger film thicknessess the self-consistent $\Delta$ would be modified by a factor of $d_0/d$. The dashed line indicates the room temperature energy scale.}
\end{figure}

To assess the prospects of these materials more realistically, we consider more closely one of the most promising candidates, namely Bi$_2$Se$_3$~\cite{TIreviews}. It has a Fermi velocity of about $5 \times 10^5$m/s which leads to an
effective fine structure constant $\alpha_0\sim 4.4$ in the
absence of dielectric screening.  Unfortunately, Bi$_2$Se$_3$ has a large dielectric constant.
To account for its influence we model the bulk of the topological insulator as
a uniform dielectric and assume that the thin film is surrounded by an extrinsic embedding dielectric.
Then, the bare intra and interlayer interactions are~\cite{Dagim,Profumo},
\be
\nu_0 V_S(q)=\frac{e^2}{\epsilon_S v_D q}\frac{2 [(\epsilon+1)e^{qd}+(\epsilon-1)e^{-qd}]}{[(\epsilon+1)^2e^{qd}-(\epsilon-1)^2e^{-qd}]},
\ee

\be
\nu_0 V_D(q)=\frac{e^2}{\epsilon_S v_D q}\frac{4  \epsilon}{[(\epsilon+1)^2e^{qd}-(\epsilon-1)^2e^{-qd}]},
\ee
where $\epsilon=\epsilon_{TI}/\epsilon_S$ is the ratio of the dielectric constant of the topological insulator ($\epsilon_{TI}$) to that of the surrounding medium ($\epsilon_S$). For very thin topological insulator films, the Dirac dispersion is modified by single-particle hybridization between top and bottom surfaces~\cite{TIthinfilms}.  This limits the smallest thicknesses of thin films for which we can hope to observe interaction induced coherence effects to a few quintuple layers ($QL$), $d \gtrsim 3QL \sim 3 {\rm nm}$. Constant gap approximation results obtained for the fixed ratio $\Delta/\epsilon_F=0.1$ and $d = 3 {\rm nm}$ are shown in Fig.~\ref{TIscreening}, by fixing the ratio to be an order of magnitude smaller than the Fermi energy we have considered the limitation that the Kosterlitz-Thouless transition poses in order to provide an estimate for the highest transition temperatures achievable in this material.

From the results of this section we can say that the prospect for observing interlayer coherence
in graphene and topological insulators is optimistic, provided that small interlayer separations
can be achieved and extrinsic substrate screening minimized.  The prospects of approaching room
temperature are more favorable for graphene than for TI thin films in spite of its larger number of
screening channels, mainly because of the possibility of achieving larger Fermi energies but
also because the TI surface state layers are separated by the TI bulk which screens interactions strongly.
The energetically remote $\pi$-bands will further favor interlayer coherence in graphene as pointed out
previously~\cite{mink,register1,register2,lozovik3} and discussed further in Sec.~\ref{remotebands}.

\section{Momentum dependent gap}\label{momentumg}

\begin{figure}
\begin{center}
\includegraphics[scale=0.8]{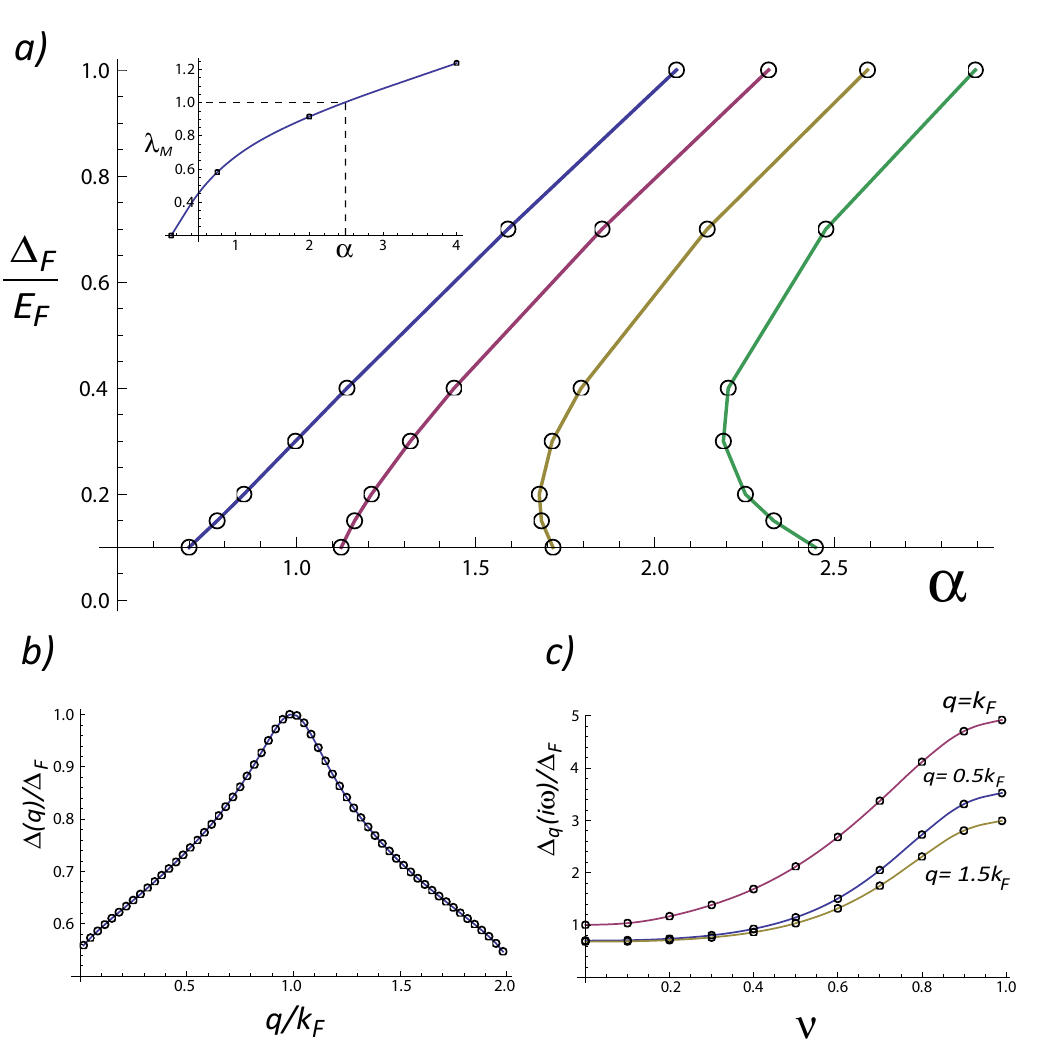}
\end{center}
\caption{\label{lineargap} a) Value of the momentum dependent excitonic gap at $q=k_F$ as a function of the effective fine structure constant, determined from the eigenvalue problem described in the text. This figure corresponds to the limit $d\rightarrow 0$ and the curves are for different Dirac cone degeneracies $N=\{1,2,3,4\}$. Notice that in graphene ($N=4$), the gap is not single valued as a function of $\alpha$ indicating that a discontinuous phase transition can occur even for $k_Fd=0$ when the momentum dependence of the gap is taken into account (compare with the results within the constant gap approximation of Fig.~\ref{spont_gap}). The inset shows the largest eigenvalue of the integral operator for $\Delta= 0.1 \epsilon_F$ as a function of $\alpha$ for $N=4$, self-consistency is achieved at an $\alpha$ for which $\lambda_M=1$. b) Momentum dependence of the gap corresponding to the same parameters used in the inset of a). c) Frequency dependence of the gap as a function of the non-linear frequency scale, $\omega/\epsilon_F\equiv \nu/(1-\nu)$, estimated by replacing the momentum dependent of gap of panel (b) in the L.H.S. of Eq.~\ref{biggapeq}. The smooth frequency-momentum variation across the fermi surface over a scale $~\Delta/v_D$ validates the consistency of the approach.}
\end{figure}

A simple way to include the momentum dependence of the gap is to cast Eq.~\eqref{biggapeq} as an eigenvalue problem. To do so, we replace the gap by its value at $\Omega=0$ and $q=k_{F}$ in the energy denominator and in the screening function while retaining full momentum dependence in the numerator of the anomalous Green's function. This procedure is justified if we assume a smooth wavevector dependence of the gap near the Fermi surface over a scale $\Delta/v_F$, and a weak frequency dependence over a scale $\epsilon_F$, below we illustrate that these two conditions are met. With these simplifications the gap equation becomes an eigenvalue problem
\begin{widetext}
\be\label{gapeq2}
\Delta(q)=\int\frac{d^2 k}{(2\pi)^2}\frac{1+\cos(\phi_{q}-\phi_{k})}{2}\int_0^{\infty}\frac{d\omega}{\pi}\frac{\Delta(k)}{\omega^2 + \Delta^2+\xi_{k}^2}
\; V^{sc}_{D}(|k - k_F \hat{q}|,i\omega) .
\ee
\end{widetext}
This equation is solved when an eigenvalue of the integral operator is equal to $1$.
By demanding that the solution is a fixed point of the integral equation,
we conclude that the physically relevant solution corresponds to the largest
eigenvalue $\lambda_M$ which must equal $1$.
By computing this eigenvalue as a function of the value of the gap at the Fermi surface, which enters the integral kernel through the energy denominator and the screened potential, we can solve for $\lambda_M(\Delta)=1$ to find the self-consistent value of the gap at the Fermi surface. In this approach $\lambda_M$ therefore plays a role similar to that played by
 $f(t)$ in the constant gap approximation.
The eigenvector corresponding to the largest eigenvalue, normalized to have a value $\Delta$ at $q=k_F$,
is the momentum dependent gap function. Compared with the constant gap approximation we find that
accounting for the momentum dependence of the gap reduces its value at the Fermi surface, although the qualitative trends towards coherence obtained within the constant gap approximation are the same, so that the conclusions of the previous section largely stand.

Note that because we include $\Delta$ in the energy denominator and in
screening, we are not solving a linearized gap equation like the one
used in BCS theory to estimate the critical temperature.
Non-linearity is encoded in the self consistency of the largest eigenvalue.
This approach is justified when the imaginary frequency dependence of the gap is
weak and gaps are smoothly varying over a width $\delta q \sim \Delta/v_D$ around the Fermi surface,
since this is the momentum region that contributes to the integral kernel.
Both assumptions are justified as illustrated in Figs.~\ref{lineargap}b) and~\ref{lineargap}c).

\section{Contributions from remote bands}\label{remotebands}

So far we have been dealing with a simplified two-band model to show that a self-consistent inclusion of the excitonic effects into the screening properties of a bilayer system considerably enhances its tendency towards interlayer coherence. Further, we used a hard momentum cut-off at $2k_F$ in the calculations. In this section we discuss what would be the effect of including the momenta beyond our $2k_F$ cutoff, and the remote Dirac bands which are not crossed by the chemical potential. Generally two competing factors appear, the enhancement of screening from the addition of available states for single particle excitations, and the increment of available phase space for coherence with a tendency to raise the gap as we will describe in this section.

Consider the full four band model equations with band diagonal pairing (see e.g. Eq.~(3.1) of ref.~\cite{lozovik2}):
\begin{widetext}
\be\label{4bandgap}
\Delta_s(q,\omega)=\sum_{s'}\int\frac{d^2 q' d\omega'}{(2\pi)^3}V^{sc}_D(q-q',i(\omega-\omega')) \; \frac{1+ss' \cos(\phi')}{2} \; \frac{\Delta^{'}_{s'}}{\omega^{'2}+\Delta^{'2}_{s'}+\xi^{'2}_{s'}},
\ee
\end{widetext}
where $s=\{+,-\}$, $\xi_s=v_D(q+sk_F)$ and the $'$ accent on $\Delta$ indicates that its arguments are $\omega',q'$.
The point we wish to make can be recognized by considering
the limit $k_F\rightarrow 0$ of this equation for which
$\Delta_+=\Delta_-\equiv\Delta$.  The gap equation for
spontaneous coherence in this limit is identical to the spin-density wave
gap equation~\cite{neutralgap} of a single neutral graphene layer, except that in the present case the interaction is
an interlayer rather than an intralayer interaction:
\begin{widetext}
\be\label{neutralgap}
\Delta(q,\omega)=\int\frac{d^2 q' d\omega'}{(2\pi)^3} \; V^{sc}_D(q-q',i(\omega-\omega')) \; \frac{\Delta^{'}}{\omega^{'2}+\Delta^{'2}+v_D^2 q^{'2}},
\ee
\end{widetext}
In the single-layer case it is known that beyond a critical interaction strength a gap will open due to coherence
between conduction and valence bands that marks the onset of a transition to a
gapped state with sub-lattice spin-density wave ordering~\cite{neutralgap}.
(This ordered state is sometimes referred to as an excitonic state, although this
terminology is not appropriate since the number of states in the valence and
conduction bands of a single layer are not separately conserved, even approximately.) 
Since the neutral system lacks an energy scale, the size of such gap is determined by ultraviolet scales
beyond the Dirac model.  Just as a spin-density wave state is a realistic
possibility in neutral single-layer graphene, an interlayer coherent state is a
realistic possibility in graphene double-layers {\em even in the absence of carriers.}
States that lie above the $2 k_{F}$ ultraviolet cut-off we have employed so
far can induce coherence on their own if interlayer interactions are strong enough,
so they must have an influence even at more moderate interaction strengths.

To illustrate their influence, we separate the integral in Eq.~\eqref{4bandgap} into two contributions,
one for momenta running up to $2 k_F$ and one for momenta running from $2 k_F$ up to a
new UV momentum $p_w$ on the Brillouin-zone scale.
For simplicity we assume that $\Delta_+=\Delta_-\equiv\Delta$; it is easy to show that this
becomes exact for the case of a pseudo-potential $V^{sc}_D(q-q',i(\omega-\omega'))$ which is independent of the angle
between $q$ and $q'$.
(Note that $\Delta_+(q=0,i\omega)=\Delta_-(q=0,i\omega)$ even without this assumption, and that the
coherence of remote bands consequently does not necessarily open up a gap at the Dirac points
of each layer.)

We approximate the large momentum contribution to the right-hand-side of the gap
equation by assuming that the gap varies as $1/q$ at large $q$ because of the
large $q$ behavior of the screened Coulomb interaction, writing
$\Delta(q,i\omega)\sim\Delta_F (k_F/q)$, where $\Delta_F$ is the value of the gap at the Fermi surface. Calling $\delta \Delta_F$ the contribution to the gap at the Fermi surface coming from UV momenta, we can write for $p_w\gg k_F$,
\be
\delta\Delta_F \approx
\int_{2 k_F}^{p_w}\frac{d^2 q}{(2\pi)^2} \int\frac{d\omega}{2\pi}V^{sc}_D(q,i\omega)\frac{\Delta_F \frac{k_F}{q}}{\omega^2+v_D^2 q^2}.
\ee

The frequency integral can be approximated by viewing it as averaging the screened potential with a non-normalized Cauchy-Lorentz distribution of width $\sim v_D q$:
\begin{widetext}
\be
\int_0^{\infty}\frac{d\omega}{\pi}\frac{V^{sc}_{D}(q,i\omega)}{\omega^2 +(v_Dq)^2}=\frac{1}{2v_D q}\langle V^{sc}_{D}(q,i\omega) \rangle_\omega \sim \frac{V^{sc}_{D}(q,\mathcal{O}(v_D q))}{2v_D q}.
\ee
\end{widetext}
In the limit  $d\rightarrow 0$, it follows from the behavior of the polarization function in neutral graphene~\cite{graphenereviews} that $V^{sc}_{D}(q,\mathcal{O}(v_D q))\rightarrow (2\pi e_\star^2)/q$, where $e_\star^2$ is a function of $\alpha,N$ which measures the effective strength of the interaction at high energies. In this way we obtain,
\be
\delta\Delta_F \approx \Delta_F \frac{\alpha_\star}{4} \equiv\Delta_F\frac{\alpha_\star}{\alpha_c},
\ee
where $\alpha_\star= e_\star^2/v_D$ and $\alpha_{c}$ is the value of $\alpha_\star$ for which
coherence would occur without carriers.
Although this line of argumentation is qualitative, two points seem unavoidable: a) that the contribution to the
right-hand side of the gap equation from UV momenta is proportional to the value of the gap near the Fermi surface, and
b) that the factor of proportionality is smaller than one unless interlayer coherence would occur even
without carriers.  The absence of a spontaneous gap in neutral single layer graphene~\cite{neutralgap_exp}
suggests that spontaneous coherence gaps should also be absent in neutral double-layers and
therefore that the factor of proportionality should be smaller than one.  The UV remote band contribution
to spontaneous coherence at finite carrier density can nevertheless be substantial. By adding the large momentum contribution to the one coming from momenta $p<2 k_F$ we can
rewrite the gap equation in the form
\begin{widetext}
\be\label{gapeq3}
\Delta_F=\frac{1}{1-\frac{\alpha_\star}{\alpha_c}}\sum_{s'}\int^{2k_F}\frac{d^2 q' d\omega'}{(2\pi)^3}V^{sc}_D(q-q',i(\omega-\omega'))\frac{1+ss'\cos(\phi')}{2}\frac{\Delta^{'}}{\omega^{'2}+\Delta^{'2}+\xi^{'2}_{s'}}.
\ee
\end{widetext}
The UV contribution can be captured by multiplying the screened interaction in the gap equations
we have used in previous sections by an enhancement factor $S \equiv(1-\alpha_\star/\alpha_c)^{-1}>1$.
This analysis could be thought of as a renormalization transformation that maps the original model defined up to momentum scales of $p_w$ into a model of momenta up to $2 k_F$ while keeping the Green's functions fixed at the expense of rescaling the interactions~\cite{Shankar}.  Of course the exact transformation would also alter the frequency/momentum dependence of the interactions.

It follows from Eq.~\eqref{Vsc} that $S V^{sc}_D(q,i\omega;\alpha,N,d)= V^{sc}_D(q,i\omega;S \alpha,N/S,d)$.
The effect of the interaction enhancement is therefore two-fold, to increase the effective fine structure constant $\alpha\rightarrow S\alpha$ and to reduce the effective degeneracy of the Dirac cones $N \rightarrow N/S$.  Both
effects support coherence and their influence could be quite substantial.  These considerations
motivate the two-band calculations in Fig.~\ref{spont_gap} for values of $\alpha$ and $N$ that do not
correspond directly to either TI's or graphene.

In the right side of Eq.~\eqref{gapeq3} the bands that cross the Fermi surface have a singular contribution
in the limit $\Delta_F\rightarrow0$ which guarantees a solution.
The remote band contribution does not on its own guarantee a solution but, when included, can only increase the estimate of the gap.  This property has already been emphasized in previous studies~\cite{mink,lozovik3}.

We do note again that the polarization functions that enter Eq.~\eqref{gapeq2}
are those from the four-band model, whereas we have employed two-band polarization functions
in our two-band model calculations.  The polarization functions of the two models coincide for momentum transfers $q \lesssim k_F$, but at higher momenta the four-band model polarizations produce stronger screening. Nevertheless,
the contribution of such momenta to the two-band gap equation is reduced by the backscattering suppression factor of Dirac models and by the decay of Coulomb interactions at larger momentum transfers.

We believe our approach captures the qualitative trends in the excitonic condensation in double-layer systems, although precise quantitative predictions would require a more elaborate numerical approach which incorporates the competition of the effects described in this section.

\section{Summary and Discussion}\label{summary}

This article addresses the possibility of achieving spontaneous coherence between the layers of a graphene double-layer system, or between the top and bottom surfaces of a topological insulator thin film at temperatures needed for its robust experimental observation, or even for practical applications.

The task is accomplished by solving for the $T=0$ energy gap $\Delta$ using a $GW$ dynamic screening approximation for interlayer interactions.  The accuracy of this approximation in single-layer graphene systems is supported by detailed comparisons~\cite{ARPESexpt} with ARPES spectra. Although we believe that the approximation we employ is valuable for the purpose of identifying the most favorable circumstances for interlayer coherence, it is not fully rigorous.  For example, it neglects the rearrangement of spectral weight due to quasiparticle renormalization factors that is commonly included in strong-coupling theories of superconductivity.  More seriously, in our opinion, it does not adequately capture the competition between coherent states, which lower interlayer interaction energies, and other strongly correlated states which minimize intralayer interaction energies.

The two double-layer systems that we focused on are distinguished by the number $N$ of Dirac fermion
flavors, by their Dirac velocities $v_{D}$, and by the dielectric screening environments that surround
the two-dimensional itinerant carriers. Spontaneous coherence in these systems can be promoted
by using gates to induce two-dimensional electron and hole gases with carrier densities that are
equal in magnitude, but opposite in sign in the two layers.  Because of the particle-hole symmetry of
two-dimensional Dirac systems, this circumstance will guarantee accurate nesting between the
electron-like and hole-like Fermi surfaces. 
Under these circumstances, it is generally thought that repulsive Coulomb interactions between electrons in different layers are guaranteed to induce spontaneous coherence at sufficiently low temperatures.  For small ratios of the excitonic gap, $\Delta$, to the carrier Fermi energy $\epsilon_F$,
mean-field theory is expected to be accurate
at finite-temperatures.  Recalling the familiar result from the BCS theory of superconductivity~\cite{BCS}, this
implies a critical temperature $k_{B} T_{c} \sim 0.5 \Delta$.  At larger values of $\Delta$ the
phase transition, which is of the Kosterlitz-Thouless type, will have a
critical temperature that saturates.  Because the Kosterlitz-Thouless temperature cannot be larger
 than \cite{hongki} $\sim 0.1 \epsilon_{F}$,
we are interested in identifying circumstances under which $\Delta \sim 0.1 \epsilon_{F}$ or larger. Under these circumstances, achieving high temperature coherence is simply a matter of increasing the carrier Fermi energy. 

When the retarded character of the screened interlayer interaction is ignored by assuming the statically screened form
of the Coulomb interaction to be valid up to frequencies of the order of the Fermi energy, it
can be shown that the gap can, at best, reach scales of $\Delta \lesssim \epsilon_{F} \exp (-4N)$~\cite{Kharitonov}, where the flavor number $N=4$ for graphene and $N=2$ for topological insulator surface states.
Since the largest achievable~\cite{graphenereviews,TIreviews} Fermi energies are smaller than $\sim 1 {\rm eV}$ in current
graphene samples and smaller than $\sim 0.1 {\rm eV}$ in current topological insulators, the largest achievable gaps
would be $\sim 10^{-7} eV$ in graphene double-layers and $\sim 10^{-3} eV$ in topological insulators if this
approximation was accurate. Our main finding, which is summarized in Fig.~\ref{spont_gap}, is that a self-consistent description of dynamical screening in the gapped phase leads to zero temperature gaps which become of the order of the Fermi energy when the layers are close together and the dimensionless coupling constant which characterizes interactions strengths $\alpha = e^2/\epsilon \hbar v_{D}$ is large.

The double-layer electron-hole pairing problem shares some features with the BCS model of electron-electron pairing in conventional superconductors. There are nevertheless important differences, one of the most prominent ones being the lack of an intermediate energy scale between the gap and the Fermi energy which in conventional superconductivity is provided by the Debye frequency. In the present case the screened interaction crosses over as a function of frequency from the statically screened values at zero frequency to the unscreened Coulomb potential at frequencies on the order of the
momentum-transfer dependent plasmon energies~\cite{lozovik2}, which vanish at large wavelengths in two-dimensional electron systems. Since the momentum transfer for scattering at the Fermi surface can take values $q\in[0,2k_F]$, the small momentum
transfer region is particularly poorly described by a statically screened interaction.  This effect is responsible for increasing the gap by orders of magnitude~\cite{lozovik2} compared to the non-zero but unobservably small values, obtained with pure static screening.~\cite{Kharitonov}.
An additional enhancement is brought about by self-consistently accounting for the effect of the gap on screening.
The gap suppresses screening only at small momenta $q\lesssim \Delta/v_D$, but this is
where the Coulomb interaction is strongest.  Therefore accounting for the decrease of screening when the gap opens
makes a sizable contribution to the right hand side of the gap equation.
These different approximations are compared directly in Fig.~\ref{spont_gap}.  The final result is that the
gap changes relatively suddenly, sometimes discontinuously, from a value that is orders of magnitude smaller than $\epsilon_{F}$ to a value larger than $0.1 \epsilon_{F}$ at a critical interaction strength which is close to the values that can be achieved by minimizing dielectric screening.

Self-consistent screening of Coulomb interaction in the presence of an excitonic gap has important consequences for the fate of the condensate as the interlayer distance is increased. In particular, we found that a first order phase transition from a large gap state to a small gap state occurs as interlayer separation increases. In the presence of disorder the smaller-gap phase would be hard to distinguish from a normal phase, thus in experiments this transition should effectively look like a first order transition between coherent (large-gap) and normal phases.

The results in this work were obtained by considering only those two-dimensional bands
which have Fermi surfaces.  In the case of topological insulator thin films there are many bulk-like bands
at higher energies.  It seems likely that the only important role of these bands is in providing additional
screening which can be approximately captured by modeling the TI bulk as a uniform dielectric.

The remote Dirac bands, that is the valence band of the n-type layer and the conduction band of the p-type layer, play a special role in establishing the interlayer coherence in graphene. Their supportive role, described in more detail in Sec.~\ref{remotebands}, can be understood from the following qualitative argument. In the two-band model the self-energy responsible for the order of the condensed state induces coherence between the conduction band of the n-type layer and the valence band of the p-type layer.  In the language of graphene's sublattice pseudospin (or real spin in the case of TIs) such a self-energy represents an interlayer coupling between antiparallel pseudospins. Because the remote band states in each layer differ from those of the bands with Fermi surfaces only by pseudospin reversal, a self-energy that couples one set of bands will also couple the other set.  The two contributions to interlayer coherence are in phase at all wavevectors.  The remote bands can have a large impact, even though they are removed energetically, because they are present over a very wide energy range.  The pseudospin reversal relationship between conduction and valence bands in graphene which leads to this property is paralleled by a real spin reversal relationship in topological insulators.  The remote band effect therefore applies equally well to that case, but over a much more limited energy range. The quantitative impact of remote bands is difficult to estimate, however, as discussed in Sec.~\ref{remotebands}. If coherence does occur in the range of interaction strengths and layer separations that is achievable experimentally, the remote band effect could be essential.

Our results indicate that reducing extrinsic screening is crucial for the observation of interlayer coherence.  Consider the case of hexagonal boron nitride (h-BN), an attractive material for building high quality graphene hetero-structures~\cite{hBN}. h-BN has a dielectric constant $\epsilon \sim 5$~\cite{hBN2}, thus, double-layer graphene fully embedded in h-BN would have an effective fine structure constant $\alpha \sim 0.4$, and our estimates suggest that the gap will only rise to values $\Delta \lesssim 10^{-5} \epsilon_F \sim 0.1$K at $\epsilon_F \sim 1$eV. On the other hand, for a heterostructure where h-BN is only used as a substrate and as a thin barrier while the top layer is surrounded by vacuum we have an effective dielectric constant $\epsilon \sim 3$, for which $\alpha \sim 0.7$ and $\Delta \lesssim 10^{-4} \epsilon_F \sim 1$K. Finally, a heterostructure where h-BN is used only as a thin barrier between the layers while the upper and lower halves are surrounded by vacuum should have an effective fine structure constant near that of graphene in vacuum $\alpha \sim 2.2$, and our results indicate that in such case the zero temperature gap will become of the order of the Fermi energy and interlayer coherence could survive to ambient temperatures.

After completing this work we became aware of a recent study which reaches similar conclusions to the ones we have presented~\cite{Lozovik_recent}.

\begin{acknowledgments}
The authors are thankful to Sanjay Banerjee, Rembert Duine, Vladimir Falko, Insun Jo, Chris Mann, Martijn Mink, Frank Register, Henk Stoof, and Emanuel Tutuc for valuable discussions. IS and AHM would like to thank the warm hospitality of KITP where part of this work was carried out. This work was supported by Welch Foundation Grant No. TBF1473, by DOE Division of Materials Sciences and Engineering Grant No. DEFG03-02ER45958, by National Science Foundation Grant No. NSF PHY05-51164, and by the NRI SWAN program.
\end{acknowledgments}

\end{document}